\documentclass[prd,tightenlines,nofootinbib,superscriptaddress]{revtex4}

\usepackage{amsfonts,amssymb,amsthm,bbm,amsmath}

\usepackage{hyperref}

\usepackage{color,psfrag}
\usepackage[dvips]{graphicx}

\newcommand{\C}{{\mathbb C}}

\newcommand{\R}{{\mathbb R}}

\newcommand{\cG}{{\mathcal G}}

\newcommand{\cM}{{\mathcal M}}

\newcommand{\cP}{{\mathcal P}}

\newcommand{\cS}{{\mathcal S}}
\newcommand{\cU}{{\mathcal U}}
\newcommand{\SU}{\mathrm{SU}}

\newcommand{\SL}{\mathrm{SL}}

\newcommand{\U}{\mathrm{U}}
\newcommand{\ISO}{\mathrm{ISO}}
\newcommand{\SB}{\mathrm{SB}}

\newcommand{\be}{\begin{equation}}
\newcommand{\ee}{\end{equation}}
\newcommand{\beq}{\begin{eqnarray}}
\newcommand{\eeq}{\end{eqnarray}}
\newcommand{\bes}{\begin{eqnarray}}
\newcommand{\ees}{\end{eqnarray}}

\newcommand{\mat} [2] {\left ( \begin{array}{#1}#2\end{array} \right ) }
\newcommand{\tabl} [2] {\begin{array} {#1} #2 \end{array}}

\newcommand{\su}{{\mathfrak su}}

\renewcommand{\u}{{\mathfrak u}}
\renewcommand{\sl}{{\mathfrak sl}}

\newcommand{\la}{\langle}
\newcommand{\ra}{\rangle}

\newcommand{\tr}{{\mathrm{Tr}}}
\newcommand{\f}{\frac}
\newcommand{\tl}{\widetilde}

\newcommand{\bz}{\overline{z}}

\def\poi#1{\{ #1 \}}

\def\nn{\nonumber}

\def\arr{\rightarrow}
\def\tN{{\widetilde{N}}}

\def\ka{\kappa}

\def\mone{^{{-1}}}

\def\tz{\widetilde{z}}
\def\eps{\epsilon}

\newcommand{\id}{\mathbb{I}}

\def\vsigma{\vec{\sigma}}
\def\ka{\kappa}

\def\UQ{{\cU_{q}(\su(2))}}
\def\ot{\otimes}

\newtheorem{theorem}{Theorem}[section]

\def\bzeta{\bar{\zeta}}
\def\vX{\vec{X}}
\def\vT{\vec{T}}
\def\tX{\widetilde{X}}
\def\tzeta{\tilde{\zeta}}
\def\bt{\bar{t}}
\def\btau{\bar{\tau}}
\def\tu{\tilde{u}}
\def\tell{\tilde{\ell}}
\def\tT{\tilde{T}}
\def\tt{\tilde{t}}
\def\ttau{\tilde{\tau}}
\def\tlambda{\tilde{\lambda}}
\def\tz{\tilde{z}}
\def\tl{\tilde{l}}
\def\btz{\bar{\tz}}

\def\tN{\tilde{N}}

\def\balpha{\bar{\alpha}}

\def\bbeta{\bar{\beta}}
\def\btt{\bar{\tt}}
\def\vt{\vartheta}

\begin{document}

\title{Deformed Spinor Networks for Loop Gravity: \\Towards Hyperbolic Twisted Geometries}

\author{{\bf Ma\"it\'e Dupuis}}\email{m2dupuis@uwaterloo.ca}
\affiliation{Department of Applied Mathematics, University of Waterloo, Waterloo, Ontario, Canada}

\author{{\bf Florian Girelli}}\email{fgirelli@uwaterloo.ca}
\affiliation{Department of Applied Mathematics, University of Waterloo, Waterloo, Ontario, Canada}

\author{{\bf Etera R. Livine}}\email{etera.livine@ens-lyon.fr}
\affiliation{Laboratoire de Physique, ENS Lyon, CNRS-UMR 5672, 46 all\'ee d'Italie, Lyon 69007, France}

\date{\today}

\begin{abstract}
In the context of a canonical quantization of general relativity, one can deform the loop gravity phase space  on a graph by replacing the $T^*\SU(2)$ phase space attached to each edge by  $\SL(2,\C)$ seen as a phase space. This deformation is supposed to encode the presence of a non-zero  cosmological constant.
Here we show how to parametrize this phase space in terms of spinor variables, thus obtaining deformed spinor networks for loop gravity, with a deformed action of the gauge group $\SU(2)$ at the vertices. These are to be formally interpreted as the  generalization of loop gravity twisted geometries to a hyperbolic curvature.

\end{abstract}

\maketitle

\section*{Introduction}


Loop quantum gravity is an approach to quantum gravity, based on a canonical quantization of general relativity formulated as a gauge field theory (typically $\SU(2)$ or $\SL(2,\C)$ depending on the precise formulation). It defines quantum states of geometry and encodes the theory's dynamics in quantum Hamiltonian constraints generating space-time diffeomorphisms. In this context, the cosmological constant $\Lambda$ is a mere coupling constant (for the 3-volume term) entering the Hamiltonian. 

There are nevertheless many claims that we could encode it in a quantum deformation of the gauge group. Such claims are backed up by rigorous analysis in 2+1-dimensional quantum gravity, for which there are clear relations between the cosmological constant and quantum group deformation (see e.g. \cite{Schroers:2011wn} for a summary). 
These studies are mostly based on the Chern-Simons reformulation of 2+1d gravity, seen from the perspectives of path integral quantization \cite{Witten1,Witten2} and Hamiltonian or combinatorial quantization \cite{Alekseev:1994pa,Buffenoir:2002tx,Meusburger:2008bs, cat}, but also on the Turaev-Viro state-sum model  \cite{Turaev:1992hq} defining a topological spinfoam path integral for 3d quantum gravity based on $\cU_{q}(\su(2))$ and believed to account for a non-vanishing cosmological constant (based on the asymptotics of the quantum-deformed 6j-symbol) \cite{Mizoguchi:1991hk,woodward}. We refer also to \cite{Freidel:2004nb} for the link between the spinfoam quantization and the combinatorial quantization and  the more recent works \cite{ours1,ours2,KauffmanBracketLQG, pranzetti} for attempts to relate the spinfoam framework and the canonical loop gravity quantization.  

The previous work \cite{HyperbolicPhaseSpace} showed how to deform at the classical level the loop gravity phase space (on a fixed graph) and analyzed in details its deformed symmetries. It was then shown in \cite{ham} that its quantization  yields to spin networks states based on the quantum gauge group $\cU_{q}(\su(2))$. The results in \cite{HyperbolicPhaseSpace}  were based on the deformation of the $T^*\SU(2)$ phase space identified as the (double cover of the) $\ISO(3)$ Poincar\'e group  to the $\SL(2,\C)$ phase space with Poisson brackets defined in terms of the classical $r$-matrix of $\sl_{2}$. The deformed loop gravity phase space on a given graph was then defined as many copies of this $\SL(2,\C)$ phase space together with a deformed action of the $\SU(2)$ gauge group on the data at the graph vertices. We further discussed the geometrical interpretation of this deformed loop gravity phase space on (3-valent) graphs within the context of  3d gravity. More precisely, we proved that a graph dressed with $\SL(2,\C)$ group elements along the edges,  provided with appropriate gauge invariant $\SU(2)$ flatness constraints, could be interpreted as defining a discrete hyperbolic surface, built from hyperbolic triangles (dual to the graph vertices) glued together consistently within the 3d one-sheet-hyperboloid (defined as set of unit time-like vectors in 4d Minkowski space-time).

Here we introduce and investigate a reformulation of this hyperbolically-deformed phase space in terms of spinor variables.
Indeed in standard loop quantum gravity (with $\Lambda=0$), there has  been a growing interest in the reformulation of the loop gravity phase space in terms of twisted geometries and spinor networks \cite{twisted1,twisted2,twisted-rev,un-spinor,spinor-lqg,spinor-holonomy}. These tools brought a clarification of the geometrical meaning of the holonomy-flux observables on a graph at the classical level and of spin network states at the quantum level. A big advantage of this formalism is a straightforward quantization of the spinor variables parametrizing the phase space and a straightforward  construction of coherent states of geometry \cite{un-spinor,spinor-coh1,spinor-coh2}. They also led to new insights on spinfoam amplitudes (for the dynamics of spin networks in loop quantum gravity) and on the associated  $3nj$-symbols of spin recoupling, providing a new light on the recursion relations and  generating functions for the amplitudes and allowing for some exact analytical evaluations \cite{Freidel:2012ji,Bonzom:2011nv,Livine:2011up,Bonzom:2012bn}. This spinorial formalism also turned out to be relevant from a purely mathematical point of view, see e.g. \cite{francesco}.

We set on introducing a similar parametrization of the $\SL(2,\C)$ phase space in terms of  spinor variables. Given an oriented graph, we consider two spinors (complex 2-vectors) on each edge, one at each end of the edge. We will reconstruct the whole $\SL(2,\C)$ phase space for these spinors, both the vector variables (the generalization of the fluxes), identified as triangular matrices, and the $\SU(2)$ holonomy along the edge. We will identify canonical spinor variables, as Darboux coordinates for the phase space, that do not transform simply under $\SU(2)$ transformations, and we will also define non-canonical spinors, that transform properly as spin-$\f12$ vectors for the fundamental representation of $\SU(2)$, from which we can directly reconstruct the vector variables as quadratic polynomials in their components.

The paper goes as follows. In a first section, we review the standard construction of the $T^*\SU(2)\sim \ISO(3)$ phase space from spinor variables. We review in the second section the definition of the $\SL(2,\C)$ phase space, parametrized in terms of triangular matrices and $\SU(2)$ group elements. The third and fourth sections introduce the  new spinor variables and  shows how to define the triangular matrices from them. The fifth section  describes  how to reconstruct the $\SU(2)$ group elements and thus how to recover the whole $\SL(2,\C)$ phase space from the spinors. The sixth section summarizes the new spinorial parametrization of the hyperbolically-deformed loop gravity phase space and the deformed action of $\SU(2)$ transformations on all the variables.
We conclude with a discussion on the possibilities opened by this formalism.

\section{Standard spinor phase space and holonomy reconstruction}

Let us consider a spinor $|\zeta \ra \, \in \C^2$ and its conjugate $\la \zeta | \, \in \C^2$,
\be
|\zeta \ra=\left( \tabl{c}{ \zeta_0 \\ \zeta_1}\right),\qquad \la \zeta |=(\bar{\zeta}_0, \, \bar{\zeta}_1),
\ee
provided with the canonical Poisson brackets:
\be
\{\zeta_0, \, \bar{\zeta}_0\}=\{ \zeta_1, \, \bar{\zeta}_1\}=-i.
\ee
We also introduce the dual spinor:
\be
|\zeta]=\mat{cc}{-\bzeta_{1} \\ \bzeta_{0}}
=\mat{cc}{0 & -1 \\ 1 & 0 } |\bzeta \ra,\qquad
[\zeta|=\mat{cc}{-\zeta_{1} \\ \zeta_{0}}\,.
\ee

We will also write $N_{A}=\zeta_{A}\bzeta_{A}$ for the modulus of the spinor components for $A=0,1$ and $N=N_{0}+N_{1}$ for their sum. It is easy to check that these generate dilatations on the complex variables:
$$
\{ N_A, \zeta_B \}=i \delta_{AB}\zeta_A\,,
\qquad
\{N_A, \bar{\zeta}_B\}=-i\delta_{AB}\bzeta_B\,.
$$

\medskip

Considering the rank-1 Hermitian matrix $X\equiv\,|\zeta\ra\la\zeta|$, one defines its projection onto the identity and the Pauli matrices $\sigma_{i}$ for $i=1..3$:
\be\label{flux}
X_{0}=\f12\tr |\zeta\ra\la\zeta| =\f12\la\zeta|\zeta\ra =\f{N_{0}+N_{1}}2=\f N2,
\qquad
\vX =\f12\tr \,\vsigma |\zeta\ra\la\zeta| =\f12\la\zeta|\vsigma|\zeta\ra\,\in\R^3\,,
\qquad
X=\left(X_{0}\id+\vX\cdot\vsigma\right)\,.
\ee
These satisfy the obvious equality $X_{0}=|\vX|$ and their Poisson brackets form a $\su(2)$ algebra:
\be
\{X_{i},X_{j}\}=\eps_{ijk}X_{k}\,,
\qquad
\{X_{0},\vX\}=0\,.
\ee
We can also decompose this in self-dual and anti-self-dual components:
$$
X=X_{0} \id +X_{3}\sigma_{3} +X_{+}\sigma_{-}+X_{-}\sigma_{+}=
\mat{cc}{(X_{0}+X_{3}) & X_{-} \\ X_{+} & (X_{0}-X_{3}) },
\qquad
X_{+}=\bzeta_{0}\zeta_{1},\quad
X_{-}=\zeta_{0}\bzeta_{1},$$
\be
\{X_3,X_\pm\}=\mp i X_{\pm}\quad
\{X_+,X_-\}=-2iX_{3}\,.
\ee
Note that $[\zeta|\vsigma|\zeta]=-\la\zeta|\vsigma|\zeta\ra$ hence the dual spinor $|\zeta]$ defines the opposite vector $-\vX$. The vector $\vX$ actually generates $\SU(2)$ transformations on the spinor:
\be
\{\vX,\,|\zeta\ra\}
=
\f i2 \vsigma\,|\zeta\ra,
\qquad
e^{\{\vec{u}\cdot\vX,\,\bullet\}}\,|\zeta\ra
=
e^{\f i2 \vec{u}\cdot\vsigma}\,|\zeta\ra,
\qquad
e^{\f i2 \vec{u}\cdot\vsigma}\in\SU(2)\,.
\ee
These finite $\SU(2)$ transformations act on the spinor $\zeta$ as 2$\times$2 matrices as in the fundamental representation while they simply induce 3d rotations on the vector $\vX$:
\be
\left.
\begin{array}{lcl}
 |\zeta\ra & \longrightarrow & g\,|\zeta\ra \\
|\zeta] & \longrightarrow & g\,|\zeta] 
\end{array}
\right)\, \leadsto  \,  X \, \longrightarrow \, g\,X\,g^{-1}\,  \leadsto \, \left( 
\begin{array}{lcl}
 X_{0} & \longrightarrow & X_{0}\\
\vX & \longrightarrow & g^{-1}\triangleright\vX \\
\end{array}\right.
\ee
where the spinor $\zeta$ and its dual both transform properly under $\SU(2)$ transformations.
We can also give the details of the infinitesimal transformations of the spinor components:
$$
\begin{array}{lcl}
\{X_{3},\zeta_{0}\}=\f i2\zeta_{0}
&\quad&
\{X_{3},\zeta_{1}\}=-\f{i}2\zeta_{1} \\
\{X_{+},\zeta_{0}\}=i\zeta_{1}
&\quad&
\{X_{+},\zeta_{1}\}=0\\
\{X_{-},\zeta_{0}\}=0
&\quad&
\{X_{-},\zeta_{1}\}=i\zeta_{0}
\end{array}
$$

\medskip

\begin{figure}
\includegraphics[scale=.6]{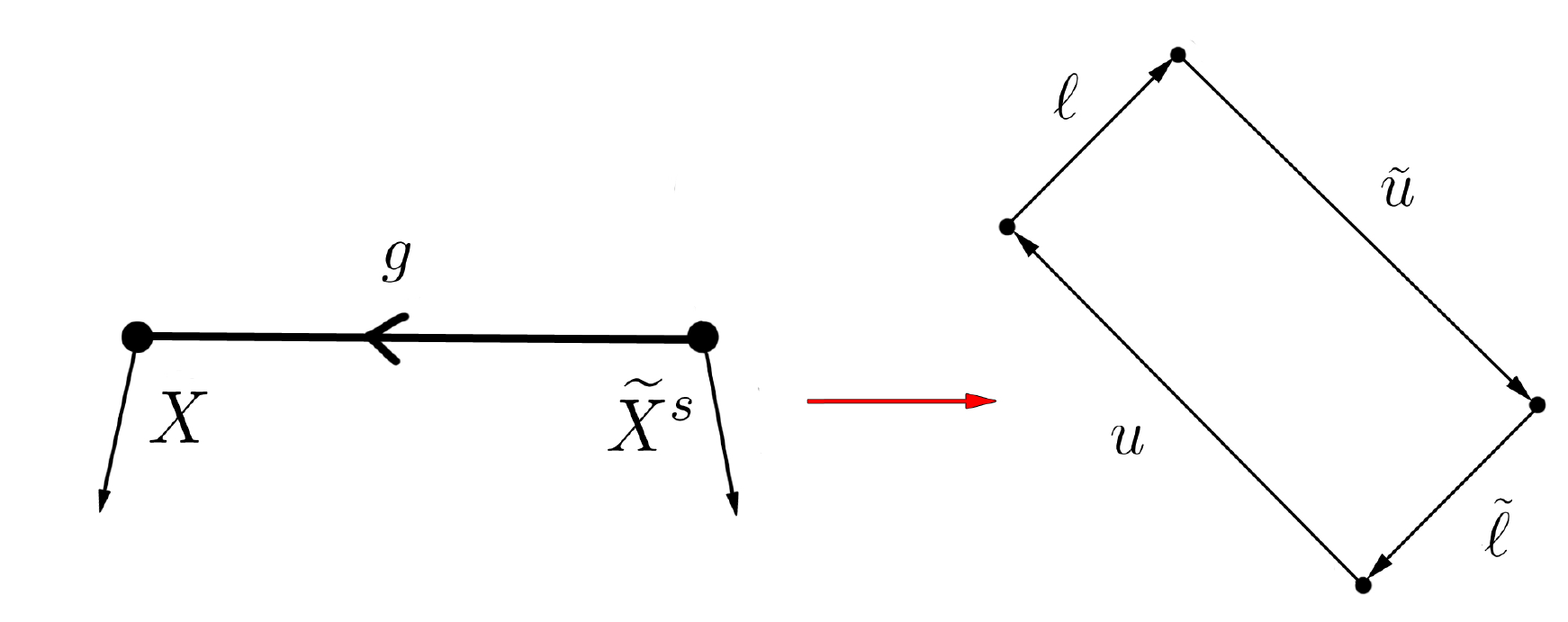}
\caption{We fatten the edges of a graph into a ribbon. The ribbon can be read clockwise as a plaquette encoding the constraint $(\tell \tu) (\ell u)= D\mone D= \id$, which encodes the equivalence of the two Iwasawa decompositions. }\label{ruban}
\end{figure}

We now consider one oriented edge of a  twisted geometry. We have two vectors at each end of the edge, $\vX$ and $\vec{\tX}$, and a $\SU(2)$ group element $g$.
We assume the standard $T^*{\SU(2)}$ Poisson brackets:
\bes
&&\{X_{i},g\}=-\f i2\,g\sigma_{i},\quad
\{\tX_{i},g\}=+\f i2\,\sigma_{i}g,\quad
\{g,g\}=0,\label{sign}\\
&&\{X_{i},X_{j}\}=\eps_{ijk}X_{k},\quad
\{\tX_{i},\tX_{j}\}=\eps_{ijk}\tX_{k}\,.
\ees
Now we further impose the condition that the holonomy $g$ sends one vector onto the other, which can be equivalent written in terms of the corresponding 2$\times$2 Hermitian matrices $X$ and $\tX$ (see Fig. \ref{ruban}):
\be
g^{-1}\triangleright \vX =\,-\,\vec{\tX},\qquad
gXg^{-1}={\tX}^{s},\qquad
\textrm{with}\quad
X=\f12(|\vX|\,\id+\vX\cdot\vsigma),\quad
\tX^{s}=\f12(|\vec{\tX}|\,\id-\vec{\tX}\cdot\vsigma)
\,.
\ee
The sign flip is important for the consistency of this condition with the Poisson brackets. Indeed it  compensates the sign difference in the brackets of $X$ and $\tX$ with the holonomy $g$ in \eqref{sign} so that the brackets of this parallel transport condition with $g$, $X$ and $\tX$ vanish.
This condition $gXg^{-1}={\tX}^{s}$ can be re-written in terms of 3d Poincar\'e transformations or equivalently as $\ISO(3)\sim\SU(2)\ltimes \R^3$ group elements represented as pairs  consisting of a rotation and a translation :
$$
(g,\vX)=(\id,\vX)(g,0)=(g,0)(\id,-\vec{\tX})=(g,-g\triangleright\vec{\tX})\,.
$$

\medskip

Noticing that the parallel transport condition $gX={\tX}^{s}g$ does not entirely fix the holonomy $g$ in terms of the two vectors $X$ and $\tX$ but leaves the possibility of an arbitrary 3d rotation around the $\vX$ axis for instance. It is possible to replace this condition by a slightly stronger one using spinor variables. This parametrization of twisted geometries  in terms of spinors allows to reconstruct both  vectors variables $X,\tX$ and $\SU(2)$ holonomies $g$ as composite observables on the  spinor phase space. These structures are called the spinor networks.

Starting with two spinors $\zeta$ and $\tzeta$ as above, we define the vectors $X$ and $\tX$ as previously:
\be
X=|\zeta\ra\la\zeta|,\quad
\tX=|\tzeta\ra\la\tzeta|\,.
\ee
We assume the norm matching condition, $\la \zeta|\zeta\ra-\la \tzeta|\tzeta\ra=0$, which implies that $\vX$ and $\vec{\tX}$ have equal norms. We then define the holonomy $g$ as the unique $\SU(2)$ group element mapping the $\C^{2}$  orthonomal basis $|\zeta\ra,|\zeta]$ to $|\tzeta],-|\tzeta\ra$:
\be
g
\,=\,
\f{|\tzeta]\la \zeta| -|\tzeta\ra[\zeta|}{\sqrt{\la \zeta|\zeta\ra\la \tzeta|\tzeta\ra}}\,\,\in\SU(2)\,,
\qquad
g\,|\zeta\ra=|\tzeta],
\quad
g\,|\zeta]=-|\tzeta\ra\,.
\ee
This properly implements the required parallel transport:
\be
g\,X\,g^{-1}=
g\,|\zeta\ra\la \zeta|\,g^{-1}
\,=\,
|\tzeta][\tzeta|
=\tX^{s}\,.
\ee
Finally, it is straightforward to check that the components of this $\SU(2)$ group element weakly commute with each other assuming the norm-matching-condition:
\be
\{g_{1},g_{2}\}\simeq0\,,
\ee
which provides us with the correct $T^{*}\SU(2)$ Poisson structure.

\medskip

Note that it is possible to avoid the sign flip of $\tX$ in this formalism. Imposing $g X =\tX g$ would require switching the sign of the brackets $\{\tX,g\}$ and $\{\tX,\tX\}$. Modifying the latter would create an asymmetry between the $X$ and $\tX$ sectors.

\section{$\SL(2,\C)$ phase space}

The $\SL(2,\C)$ phase space is defined from the $\SL(2,\C)$ group element $D$ provided with the following Poisson bracket:
\be\label{def1}
\{D_{1},D_{2}\}=-rD_{1}D_{2}-D_{1}D_{2}r^\dagger\,,
\ee
with the standard convention $D_{1}=D\otimes \id$ and $D_{2}=\id\otimes D$ and the classical $r$-matrix:
\be
r= \f{ \ka}{4}\,\sum_{i} \tau_i \ot \sigma_i =\f{i\ka}{4}\,\mat{cccc}{1 &0&0&0 \\ 0 &-1 &0 &0 \\ 0& 4 & -1 & 0 \\ 0&0&0&1}\,,
\ee
in terms of the Pauli matrices $\sigma_{i}$ and $\tau^i=i (\sigma_i - \f 12 [\sigma_3, \sigma_i])= ( i\sigma_i + \epsilon_{3i}^k \sigma_k)$. We are using here the usual notation for the tensor product of two 2$\times$2 matrices as a 4$\times$4 matrix:
$$
A_{1}B_{2}=A\otimes B =\mat{c|c}{A_{11} B & A_{12}B \\ \hline A_{21} B & A_{22}B}\,.
$$
The key to our analysis of the phase space structure  is the (left) Iwasawa decomposition $\SL(2,\C)=\SB(2,\C)\bowtie\SU(2)$:
\be
D=\ell u,
\qquad
\ell \in\SB(2,\C),\quad
u\in\SU(2)\,.
\ee
The $\SL(2,\C)$ bracket \eqref{def1} reads in this parametrization:
\be
\{\ell_{1},\ell_{2}\}=-[r,\ell_{1}\ell_{2}],\quad
\{u_{1},u_{2}\}=[r^\dagger,u_{1}u_{2}],\quad
\{\ell_{1},u_{2}\}=-\ell_{1}ru_{2},\quad
\{u_{1},\ell_{2}\}=-\ell_{2}r^\dagger u_{1}\,.
\ee
To get the full symplectic structure, one also computes the brackets with the complex conjugate variables $\ell^\dagger$. For more details, the reader can refer to \cite{HyperbolicPhaseSpace}. Defining $l={\ell^\dagger}{}^{-1}$, we get:
\be
\{l_{1},\ell_{2}\}=-[r^\dagger,l_{1}\ell_{2}]\,.
\ee
This formula contains all the brackets between  the components of $\ell$ and their complex conjugate. More precisely, we explicitly parametrize the triangular matrix as:
\be
\ell=\left( \tabl{cc}{\lambda & 0\\ z & \lambda^{-1} }\right) \in \SB(2, \C)\,,
\qquad
\lambda\in\R_{+},\quad z\in\C\,,
\ee
and compute the Poisson algebra:
\be
\{\lambda,z\}= \frac{i\kappa}{2}\lambda z,
\qquad
\{\lambda,\bar{z}\}= - \frac{i\kappa}{2}\lambda\bar{z},
\qquad
\{z,\bz\}= {i\kappa}\left( \lambda^2- \lambda^{-2} \right)\,.
\ee
Let us underline that these brackets are invariant under the exchange $\ell \longleftrightarrow \bar{\ell}^{-1}$, or explicitly $(\lambda,z) \longleftrightarrow (\lambda^{-1},\bz) $.

\medskip

Here we have used the left Iwasawa decomposition. We will also need the right Iwasawa decomposition, which corresponds to the left Iwasawa decomposition for the inverse of the $\SL(2,\C)$ group element $D$:
\be
D=\tu^{-1}\tell^{-1},\qquad
D^{-1}=\tell\tu\,.
\ee
We use a slightly different convention from the one previously used in \cite{HyperbolicPhaseSpace}, to account for the sign flip.
%
%
The Poisson brackets for the components of the inverse matrix are easy to compute:
\be
\{D^{-1}_{1},D^{-1}_{2}\}=-r^\dagger D^{-1}_{1}D^{-1}_{2}-D^{-1}_{1}D^{-1}_{2}r\,,
\ee
which amounts in the end to switching the classical r-matrix with its Hermitian conjugate.
This leads to the following brackets between the triangular matrix and the $\SU(2)$ holonomy:
\be
\{\tell_{1},\tell_{2}\}=-[r^\dagger,\tell_{1}\tell_{2}],\quad
\{\tu_{1},\tu_{2}\}=[r,\tu_{1}\tu_{2}],\quad
\{\tell_{1},\tu_{2}\}=-\tell_{1}r^\dagger\tu_{2},\quad
\{\tu_{1},\tell_{2}\}=-\tell_{2}r\tu_{1} \,.
\ee
We complete this algebra with the bracket with the complex conjugate of the triangular matrix, $\tl={\tell^\dagger}{}^{-1}$:
\be
\{\tell_{1},\tl_{2}\}=-[r,\tell_{1}\tl_{2}]\,.
\ee
Parametrizing the triangular matrix $\tell$ as before:
$$
\tell=\left( \tabl{cc}{\tlambda & 0\\ \tz & \tlambda^{-1} }\right),
$$
we get exactly the same brackets for the tilded sector as for the original sector: 
\be\label{p1}
\{\tlambda,\tz\}= \frac{i\kappa}{2}\tlambda \tz,
\qquad
\{\tlambda,\bar{\tz}\}= -\frac{i\kappa}{2}\tlambda\bar{\tz},
\qquad
\{\tz,\btz\}= {i\kappa}\left( \tlambda^2- \tlambda^{-2} \right)\,,
\ee
We can also calculate the Poisson brackets between the two different decompositions. 
\beq\label{cross bis}
\poi{\ell_1,\tell_2}=\poi{u_1,\tu_2}=0, \quad \poi{\tu_1,\ell_2}=\tu_1r^\dagger\ell_2, \quad \poi{\tell_1,u_2}=u_2r\tell_1.
\eeq

Our goal will be first to reproduce the Poisson brackets \eqref{p1} from some spinor variables and then to reconstruct the whole $\SL(2,\C)$ phase space, with in particular the $\SU(2)$ group elements $u$ and $\tu$, from those spinors.

\section{New spinor variables and deformed action of rotations}

We now define $\kappa$-deformed spinors $|\zeta^\kappa\ra$, $\la \zeta^\kappa |$ with components 
\be\label{ho def}
\zeta^\kappa_A
\equiv\zeta_A\sqrt{\f{2\sinh(\f{\kappa}{2}N_A)}{N_A}}
\,,
\quad
\bar{\zeta}^\kappa_A
\equiv \overline{\zeta^\kappa_A}
=\bzeta_A\sqrt{\f{2\sinh(\f{\kappa}{2}N_A)}{N_A}}\,.
\ee
We do not change the definition of the $N$'s, keeping $N_A^\kappa\equiv N_A=\zeta_{A}\bzeta_{A}$ for both $A=0,1$, but the norm $\zeta_A^\kappa \bar{\zeta}_A^\kappa$ of the deformed spinor has a $N_{A}$-dependent factor:
\be
\zeta_A^\kappa \bar{\zeta}_A^\kappa
\,=\,
2\sinh(\f{\kappa N_A}{2})
\,=\,
e^{\f{\kappa N_{A}}2}-e^{\f{-\kappa N_{A}}2}\,,
\ee
$$
\la \zeta^\kappa |\zeta^\kappa\ra
\,=\,
e^{\f{\ka N}2}+e^{\f{-\ka N}2}-e^{\f{\kappa(N_{0}-N_{1})}2}-e^{\f{-\kappa(N_{0}-N_{1})}2}
\,.
$$
%
These $\kappa$-deformed spinors satisfy simple Poisson brackets among themselves
\be \label{PoissonSpinorD}
\{\zeta_A^\kappa, \bar{\zeta}^\kappa_B\}= -i \,\delta_{AB}\, \kappa \cosh(\f{\kappa N_A}{2}),
\quad
\{N_{A}, \zeta_B^\kappa\}=i \,\delta_{AB} \,\zeta^\kappa_A,
\quad
\{ N_{A},\bar{\zeta}_B^\kappa\}=-i \,\delta_{AB}\, \bar{\zeta}^\kappa_A\,.
\ee
We check the limit $\ka\arr 0^{+}$ when we should recover the undeformed quantities and we get as expected at leading order in the deformation parameter:
\be
\zeta_A^\kappa \sim \sqrt{\ka}\, \zeta_{A},
\quad
\zeta_A^\kappa \bar{\zeta}_A^\kappa\sim \kappa N_{A},
\quad
\{\zeta_A^\kappa, \bar{\zeta}^\kappa_B\}\sim  -i \,\delta_{AB}\, \kappa\,.
\ee
Then, from these $\kappa$-deformed spinors, we build the lower triangular matrix $\ell\in\SB(2, \C)$ with
\be
\ell=\left( \tabl{cc}{\lambda & 0\\ z & \lambda^{-1} }\right) \in \SB(2, \C),\qquad
\lambda= \exp(-\f{\kappa}{4}(N_0-N_1)), \quad z=\bar{\zeta}_0^\kappa \zeta_1^\kappa.
\ee
It is easy to check that they satisfy the expected Poisson brackets,
\be\label{z}
\{\lambda,z\}= \frac{i\kappa}{2}\lambda z,
\qquad
\{\lambda,\bar{z}\}= - \frac{i\kappa}{2}\lambda\bar{z},
\qquad
\{z,\bz\}= {i\kappa}\left( \lambda^2- \lambda^{-2} \right).
\ee
Let us point out that the components of the triangular matrix $\ell$ all commute with $N=N_{0}+N_{1}$:
\be
\{N,\ell\}=\{N,\lambda\}=\{N,z\}=0\,,
\ee
which simply expresses invariance of $\lambda$ and $z$ under $\U(1)$ phase transformation of the spinors $|\zeta^{\ka}\ra\,\rightarrow\,e^{i\theta}\,|\zeta^{\ka}\ra$. An important remark to underline is that the parameter $z$ is not holomorphic in the spinors $\zeta^{\ka}$ and that the notion of holomorphicity depends here on the considered variables (for instance, $\zeta$ or $\zeta^{\ka}$ or $\ell$).


\medskip

Up to now, we have merely introduced a re-parametrization of the same phase space as in the undeformed case. The big difference will come in the action of the  $\SU(2)$ transformations. As derived in \cite{HyperbolicPhaseSpace}, the infinitesimal variations generated by the (left) $\SU(2)$ rotations on the $\SL(2,\C)$ phase space are given by the following re-scaled Poisson brackets with an arbitrary observable $f$:
\be
\label{su2transfneq1}
\delta_{\eps} f = -\lambda^{-2} \kappa^{-1} \{2 \epsilon_z \lambda^{2} + \epsilon_- \lambda z +\epsilon_+ \lambda \bar{z}, f\}\,.
\ee

This comes from finite $\SU(2)$ transformations on triangular matrices:
\be
\ell
\quad\overset{v\in\SU(2)}{\longrightarrow}\quad
\ell^{(v)}=v\ell v'^{-1}\,,
\ee
where $v'\in\SU(2)$ is a priori different from $v$ and compensates for the fact that triangular matrices are not stable under conjugation by $\SU(2)$ group elements. For infinitesimal transformations, these group elements read:
\be
\label{veps}
v\underset{v\sim\id}\sim
\id + i \eps\cdot\vsigma
=
\id + i \mat{cc}{\eps_{z} & \eps_{-} \\ \eps_{+} & \eps_{z}}\,,
\qquad
v'\underset{v\sim\id}\sim
\id + i \eps'\cdot\vsigma
\quad\textrm{with}\quad
\left|
\begin{array}{rcl}
\eps'_{\pm}&=&\lambda^{-2}\eps_{\pm} \\
\eps'_{z} &=& \eps_{z} +\f12(\lambda^{-1}z\eps_{-}+\lambda^{-1}\bz\eps_{+})
\end{array}
\right.
\ee
We look for a new spinor $t^{1/2}$, or in short $t$, that transforms covariantly under $\SU(2)$, that is that has the correct  infinitesimal variations under left $\SU(2)$ rotations as a spin-$\f12$ vector:
\be
|t\ra=\mat{c}{t_{0}\\ t_{1}}
\quad\overset{v\in\SU(2)}{\longrightarrow}\quad
v\,|t\ra
\underset{v\sim\id}\sim
|t\ra+i\,\mat{c}{\eps_{z}t_{0}+\eps_{-}t_{1}\\ \eps_{+}t_{0}+\eps_{z}t_{1}}\,.
\ee
This is translated in the following required Poisson bracket identities:
\be
\begin{array}{lcl}
-\lambda^{-2} \kappa^{-1} \{  \lambda^{2} , t_{0} \}=\f {i}2 t_{0}
&\quad&
-\lambda^{-2} \kappa^{-1} \{  \lambda^{2} , t_{1} \}=-\f {i}2t_{1} \\
-\lambda^{-2} \kappa^{-1} \{ \lambda z , t_{0} \}= it_{1}
&\quad&
-\lambda^{-2} \kappa^{-1} \{ \lambda z , t_{1} \}= 0 \\
-\lambda^{-2} \kappa^{-1} \{ \lambda \bz , t_{0} \}= 0
&\quad&
-\lambda^{-2} \kappa^{-1} \{ \lambda \bz , t_{1} \}= it_{0}
\end{array}
\ee
One can solve these equations explicitly and one finds two independent spinor solutions, which are simply a rescaled $\zeta^{\kappa}$ spinor and its dual spinor:
\be
|t\ra = \mat{c}{ e^{\f{\ka N_{1}}{4}}\zeta^{\ka}_{0} \\ -e^{-\f{\ka N_{0}}{4}}\zeta^{\ka}_{1} },
\qquad
|t]=\mat{c}{-\bt_{1}\\ \bt_{0}}
= \mat{c}{ e^{-\f{\ka N_{0}}{4}}\bzeta^{\ka}_{1} \\e^{\f{\ka N_{1}}{4}}\bzeta^{\ka}_{0} }.
\ee
These new spinor components satisfy the following Poisson brackets:
\bes\label{t com}
&&\{t_{0},t_{1}\}=\f{i\ka}{2}t_{0}t_{1},\quad
\{\bt_{0},\bt_{1}\}=-\f{i\ka}{2}\bt_{0}\bt_{1},\quad
\{t_{0},\bt_{1}\}=\{\bt_{0},t_{1}\}=0,\nn
\\
&&\{\bt_{0},t_{0}\}=i\ka(-\f{1}{2}|t_{0}|^{2} + e^{\f{\ka}2 N}),
\quad
\{\bt_{1},t_{1}\}=i\ka(\f{1}{2}|t_{1}|^{2} + e^{-\f{\ka}2 N})\,.
\ees
Let us underline that this algebra does not completely close since there are the $e^{\pm\f{ \ka}2 N}$ terms\footnote{We can obtain a closed algebra if we rescale our spinors $t_A$, $\bt_A$ by $e^{-\f\ka4 N}$. Such rescaled components are then the exact classical analogue of the quantum spinor variables used  in \cite{ours2}, provided we identify $t_0\equiv t_-$, $t_1\equiv t_+$ and so on.}.
We compute the new rank-one Hermitian matrix $|t\ra\la t|$ and compare it to the matrix $T=\ell \ell^{\dagger}$:
\be
|t\ra\la t|
\,=\,
\mat{cc}{|t_{0}|^{2} & t_{0}\bt_{1} \\ \bt_{0}t_{1} & |t_{1}|^{2}}
\,=\,
\mat{cc}{e^{\f{\ka N}2}-\lambda^{2} & -\lambda \bz\\ -\lambda z & e^{\f{\ka N}2}-(\lambda^{-2}+|z|^{2})}
\,=\,
e^{\f{\ka N}2}\id -\ell\ell^{\dagger}\,.
\ee
This means that the two spinors, $|t\ra$ and its dual $|t]$, are the two eigenvectors of the Hermitian matrix $T=\ell\ell^{\dagger}$.
We usually project the $T$ matrix onto the Pauli matrices to get the $\SU(2)$-covariant 3-vector $\vT$ as follows:
\be
T_{0}=\f1{2\ka}\tr \ell\ell^{\dagger}=\f1{2\ka}(\lambda^{2}+\lambda^{-2}+|z|^{2})=\f1\ka\cosh\f{\ka N}2,
\ee
$$
\vT=\f1{2\ka}\tr \ell\ell^{\dagger}\vsigma,\quad
T_{3}=\f1{2\ka}(\lambda^{2}-\lambda^{-2}-|z|^{2}),\quad
T_{+}=\f1\ka\,\lambda z,\quad T_{-}=\f1\ka\,\lambda \bz\,.
$$
$$
T=\ka\,(T_{0}\,\id+\vT\cdot\vsigma)\,.
$$
Moreover, due to the fact that $\ell$ belongs to $\SL(2,\C)$ and that its determinant is normalized to one, we have a very simple expression for the inverse of the $T$-matrix:
\be
\ka^2(T_{0}^2-\vT^2)=\det T =\det\ell\ell^{\dagger}=1
\qquad\Rightarrow\qquad
T^{-1}=\ka\,(T_{0}\,\id-\vT\cdot\vsigma)=T^s\,,
\ee
which we recognize as simply switching the sign of the 3-vector $\vT$ and thus implementing the orientation flip introduced in the undeformed case.
With these conventions, we can express the components of $|t\ra\la t|$ in terms of the vector $\vT$:
\be
\la t|t\ra=2e^{\f{\ka N}2}-2\ka T_{0}
=2\sinh\f{\ka N}2,\quad
\la t|\vsigma|t\ra=-2\ka\vT\,,
\ee
We summarize the relation between the spinor $t$ and the matrix $T$ by the formulas:
\be
\begin{array}{lclcl}
|t\ra\la t|
&=&
e^{\f{\ka N}2}\id - T
&=&
-e^{\f{-\ka N}2}\id + T^{-1} \\
|t][ t|
&=&
e^{\f{\ka N}2}\id - T^{-1}
&=&
-e^{\f{-\ka N}2}\id + T
\end{array}
\ee
All these objects all transform as follows under $\SU(2)$ transformations:
\be
\ell \longrightarrow  v\ell v'^{-1}\,\leadsto\, 
\left( \begin{array}{rcl}
T&\longrightarrow & vTv^{-1}\\
|t\ra&\longrightarrow & v\,|t\ra 
\end{array}\right.
 \, \leadsto \,
\left( \begin{array}{rcl}
T_{0}&\longrightarrow & T_{0}\\
\vT &\longrightarrow & v^{-1}\triangleright\vT
\end{array}\right.
\ee
The 3-vector $\vec T$ could be seen as the analogue of the flux $\vec X$ in the flat case, given in \eqref{flux}.  Indeed, the vector $\vec T$ appears naturally in the analysis of the deformed phase space in \cite{HyperbolicPhaseSpace}. It allows to recover some  information about the discrete geometry, such as the hyperbolic cosine law  or the flatness constraint \cite{HyperbolicPhaseSpace}. However,  it is not sufficient, we also need another vector $\vec T^{op}$ to recover this  information. $\vec T^{op}$ is a vector that transforms under the deformed $\SU(2)$ transformations $v'$ as given in \eqref{veps}. To reconstruct $\vec T^{op}$, we need therefore to introduce another type of spinor,  a "braided-covariant"  spinor $|\tau\ra$, which transforms with $v'$ and not $v$ under the $\SU(2)$ action:
\be
|\tau\ra=\ell^{-1}\,|t\ra
=
\mat{cc}{\lambda^{-1} & 0\\ -z & \lambda }\,\mat{c}{ e^{\f{\ka N_{1}}{4}}\zeta^{\ka}_{0} \\ -e^{\f{-\ka N_{0}}{4}}\zeta^{\ka}_{1} }
=
e^{\f{\ka N_{0}}{4}}\mat{c}{ \zeta^{\ka}_{0} \\ -e^{\f{\ka N}{4}}\zeta^{\ka}_{1} }\,,
\ee
which satisfies the new identity in terms of $T^{op}=\ell^{\dagger}\ell$:
\bes
&&|\tau\ra\la \tau|
=e^{\f{\ka N}2} (\ell^{\dagger}\ell)^{-1} - \id
= e^{\f{\ka N}2} (2\cosh\f{\ka N}2\,\id- \ell^{\dagger}\ell) -\id
= e^{\f{\ka N}2} (e^{\f{\ka N}2}\id - \ell^{\dagger}\ell)\,,
\\
&& \la \tau|\tau\ra
=e^{\f{\ka N}2}\tr(e^{\f{\ka N}2}\id - \ell^{\dagger}\ell)
=e^{\f{\ka N}2}\la t|t\ra
=2e^{\f{\ka N}2}\sinh\f{\ka N}2
=e^{{\ka N}}-1
\ees
This allows to decompose this matrix and express it in terms of the 3-vector $\vT^{op}$:
\be
T_{0}^{op}=\f1{2\ka}\tr\ell^\dagger\ell=T_{0},\qquad
\vT^{op}=\f1{2\ka}\tr\ell^\dagger\ell\vsigma\,, \quad 
\la \tau|\vsigma|\tau\ra=\,2\ka\,e^{\f{\ka N}2}\,\vT^{op}\,,
\ee
And we give all the relations:
\be
\begin{array}{lclcl}
e^{\f{-\ka N}2}\,|\tau\ra\la \tau|
&=&
e^{\f{\ka N}2}\id - T^{op}
&=&
-e^{\f{-\ka N}2}\id + T^{op}{}^{-1} \\
e^{\f{-\ka N}2}\,|\tau][ \tau|
&=&
e^{\f{\ka N}2}\id - T^{op}{}^{-1}
&=&
-e^{\f{-\ka N}2}\id + T^{op}
\end{array}
\ee
Let us point out the useful identity relating the dual of both covariant and braided-covariant spinors:
\be
\ell^{-1}\,|t]
\,=\,
e^{-\f{\ka N}2}\,|\tau]\,.
\ee
This new spinor clearly transforms under $v'$:
\be
|\tau\ra=\ell^{-1}\,|t\ra
\quad\overset{v\in\SU(2)}{\longrightarrow}\quad
(v\ell v'^{-1})^{-1}\,(v\,|t\ra)=v'\,\ell^{-1}\,|t\ra=v'\,|\tau\ra\,,
\ee
and one can check that we recover the correct infinitesimal variations as prescribed by the expression of the modified parameter $\eps'$ given in \eqref{veps}.
%

\medskip

Finally, let us conclude this section with the brackets between the components of the spinor $\tau$ and its complex conjugate:
\bes\label{tau com}
&&\{\tau_{0},\tau_{1}\}=-\f{i\ka}{2}\tau_{0}\tau_{1},\qquad
\{\btau_{0},\btau_{1}\}=\f{i\ka}{2}\btau_{0}\btau_{1},\qquad
\{\tau_{0},\btau_{1}\}=-\f{i\ka}{2}\tau_{0}\btau_{1},\qquad
\{\btau_{0},\tau_{1}\}=+\f{i\ka}{2}\btau_{0}\tau_{1},\nn
\\
&& \{\btau_{0},\tau_{0}\}={i\ka}\,e^{{\ka} N_{0}}={i\ka}\,(1+|\tau_{0}|^2),
\qquad
\{\btau_{1},\tau_{1}\}={i\ka}\,e^{{\ka} N}={i\ka}\,(1+|\tau_{0}|^2+|\tau_{1}|^2)\,.
\ees

\section{The tilded sector - An exact copy of the straight sector}

We repeat the same spinor construction for the tilded sector $\tell$ since the Poisson algebra of $\tlambda$ and $\tz$ is exactly the same as the one of $\lambda$ and $z$.
We thus introduce a new spinor $\tzeta$ and its $\ka$-deformed version $\tzeta^{\ka}$, from which we define the components of the new tilded triangular matrix $\tell$:
\be
\tlambda= \exp(-\f{\kappa}{4}(\tN_0-\tN_1)), \quad \tz=\bar{\tzeta}_0^\kappa {\tzeta}_1^\kappa.
\ee
These satisfy the same Poisson brackets as in \eqref{z}:
\be \label{z tilde}
\{\tlambda,\tz\}= \frac{i\kappa}{2}\tlambda \tz,
\qquad
\{\tlambda,\bar{\tz}\}= -\frac{i\kappa}{2}\tlambda\bar{\tz},
\qquad
\{\tz,\btz\}= {i\kappa}\left( \tlambda^2- \tlambda^{-2} \right)\,.
\ee
We define the new spinor $\tt$ as:
\be
|\tt\ra = \mat{c}{ e^{\f{\ka \tN_{1}}{4}}\tzeta^{\ka}_{0} \\ -e^{\f{-\ka \tN_{0}}{4}}\tzeta^{\ka}_{1} },\qquad
|\tt\ra\la \tt|
\,=\,
\mat{cc}{e^{\f{\ka \tN}2}-\tlambda^{2} & -\tlambda \bar{\tz}\\ -\tlambda \tz & e^{\f{\ka \tN}2}-(\tlambda^{-2}+|\tz|^{2})}
\,=\,
e^{\f{\ka \tN}2}\id -\tell\tell^{\dagger}
\,=\,
e^{\f{\ka \tN}2}\id -\tT
\,.
\ee
As before, all these objects transform covariantly under the suitable $\SU(2)$ transformations, that is for $w\in\SU(2)$:
\be
\tell \longrightarrow w\tell w'^{-1},\quad
|\tt\ra \longrightarrow w\,|\tt\ra,\quad
|\tt\ra\la \tt| \longrightarrow w\,|\tt\ra\la \tt|w^{-1}
\tT \longrightarrow w\,\tT w^{-1}
\,.
\ee
We also introduce the ``braided-covariant'' spinor $\ttau$,
\be
|\ttau\ra=\tell^{-1}\,|\tt\ra
=
\mat{cc}{\tlambda^{-1} & 0\\ -\tz & \tlambda }\,\mat{c}{ e^{\f{\ka \tN_{1}}{4}}\tzeta^{\ka}_{0} \\ -e^{\f{-\ka \tN_{0}}{4}}\tzeta^{\ka}_{1} }
=
e^{\f{\ka \tN_{0}}{4}}\mat{c}{ \tzeta^{\ka}_{0} \\ -e^{\f{\ka N}{4}}\tzeta^{\ka}_{1} }\,,
\ee
which transforms as $|\ttau\ra\longrightarrow\,w'\,|\ttau\ra$ under the same $\SU(2)$ transformations. It allows to recover the vector and matrix $\tT^{op}$:
\be
|\ttau\ra\la \ttau|
=e^{\f{\ka \tN}2} (\tell^{\dagger}\tell)^{-1} - \id
= e^{\f{\ka \tN}2} (e^{\f{\ka \tN}2}\id - \tell^{\dagger}\tell)= e^{\f{\ka \tN}2} (e^{\f{\ka \tN}2}\id -\tT^{op})\,.
\ee
The Poisson commutation relations between the different spinor components are essentially the same as in \eqref{t com} and \eqref{tau com}, putting some tilde everywhere. We can also consider the cross terms, that is the Poisson commutation relations between the spinors and their tilde alter ego. 
\beq\label{cross}
\poi{t_A,\tilde t_B}=\poi{t_A,\tilde \tau_B}= \poi{\tilde t_A,\tau_B}=0.
\eeq
Taking all these to be zero  ensures the commutation relations $\{\ell_1,\tilde \ell_2\}=0$, as obtained in \eqref{cross bis}.

\begin{figure}
\includegraphics[scale=.5]{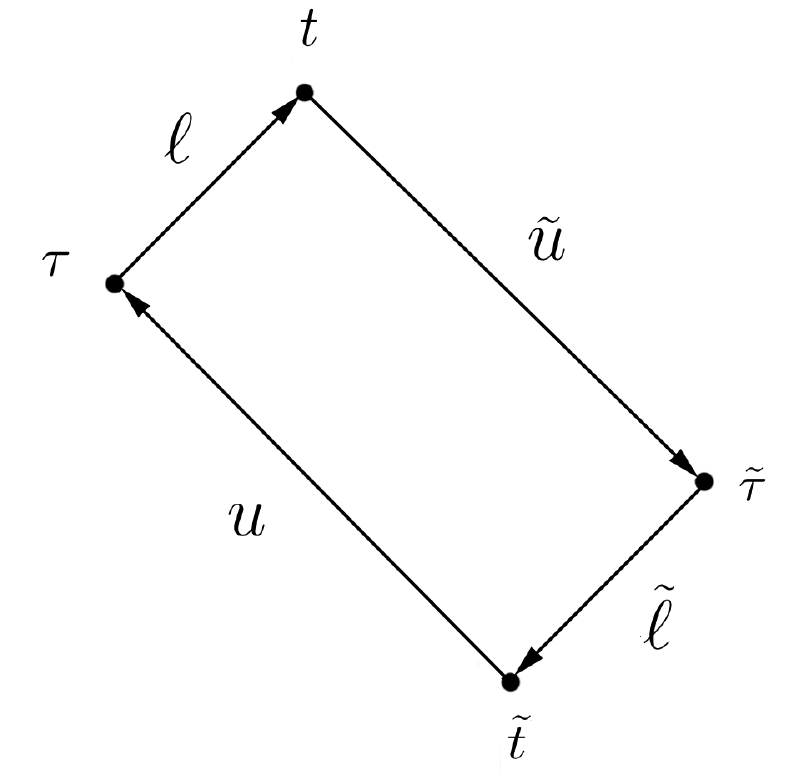}
\caption{The spinors live at the vertices of the ribbon. We can read off the mappings between spinors, $|t\ra\,=\,\ell\,|\tau\ra$ and $|\tt\ra\,=\,\tell\,|\ttau\ra$ respectively for the straight and tilded sectors separately, and $|\tau\ra\,\propto\,u\,|\tt]$ and $|\ttau\ra\,\propto\,\tu\,|t]$ for the $\SU(2)$ holonomies relating the two sectors. Notice that we have to take the dual spinors, which we haven't represented in the diagram.
This graphical representation is consistent with the realization of the symmetries (left rotation in this case) as described in Fig. \ref{leftgauge}. 
}\label{spinor-ribbon}
\end{figure}

\section{q-Deformed holonomy reconstruction}

Now starting from $\ell$ and $\tell$, i.e the 3-vectors $\vT$ and $\vec{\tT}$, we would like to reconstruct the $\SU(2)$ holonomies $u$ and $\tu$. However, as in the undeformed case, it is impossible to fully reconstruct the holonomy $g$ from the two 3-vectors $X$ and $\tX$. Indeed, we have an undetermined phase, corresponding to a free rotation around the $\vX$ axis. Nevertheless, using the spinor variables  freezes that extra phase and we can explicitly reconstruct the whole holonomy $g$ from $z$ and $\tz$ as shown in section I. Similarly here, we will use the spinors $t$ and $\tt$, from which we have already shown how to reconstruct the triangular matrices  $\ell$ and $\tell$, and we will see below how to derive the $\SU(2)$ holonomies $u$, $\tu$ and thus ultimately the $\SL(2,\C)$ group element $D$.

More precisely, from the triangular matrices $\ell$ and $\tell$, one builds the following Hermitian matrices:
$$
T=\ell\ell^{\dagger},\quad
\tT=\tell\tell^{\dagger},\quad
T^{op}=\ell^{\dagger}\ell,\quad
\tT^{op}=\tell^{\dagger}\tell\,,
$$
whose projection on the Pauli matrices define the 3-vectors  $\vT$ and $\vec{\tT}$, that live on the two ends of one edge.
The fact that $\ell$ and $\tell$ both come from the same $\SL(2,\C)$ group element $D$ decomposed following the left and right Iwasawa decomposition as $D=\ell u = \tu^{-1}\tell^{-1}$ with $u,\tu\in\SU(2)$ implies that  those 3-vectors are related by the $\SU(2)$ holonomies (obtained from the expression of $DD^{\dagger}$ and  $D^{\dagger}D$ ):
\be
\left|\begin{array}{l}
T=\tu^{-1}\tT^{op}{}^{-1}\tu \\
\vT\,=\,-\,\tu\triangleright \vec{\tT}^{op}
\end{array}\right.\,\,,
\qquad\qquad
\left|\begin{array}{l}
u^{-1}T^{op}u=\tT^{-1} \\
u\triangleright\vT^{op}\,=\,-\, \vec{\tT}
\end{array}\right.\,\,.
\ee

We now consider both left and right $\SU(2)$ transformations:
$$
D\longrightarrow vDw^{-1}=v\ell uw^{-1}=v\tu^{-1} \tell^{-1} w^{-1}\,.
$$
The left transformations read:
$$
D\longrightarrow vD=\,\,
\left\{
\begin{array}{l}
=v\ell u = (v\ell v'^{-1})(v'u) \\
=v\tu^{-1}\tell^{-1}=\left(\tu v^{-1}\right)^{-1}\,\tell^{-1}
\end{array}\right.
$$
They affect $\ell$ but not $\tell$. On the other hand, right transformations read:
$$
D\longrightarrow Dw^{-1}=\,\,
\left\{
\begin{array}{l}
= \ell (uw^{-1}) \\
=\tu^{-1} \tell^{-1} w^{-1} =\left(w'\tu\right)^{-1}\,\left(w\tell w'^{-1}\right)^{-1}
\end{array}\right.
$$
and likewise affect only $\tell$ while leaving $\ell$ unchanged.
\begin{figure}
\includegraphics[scale=.4]{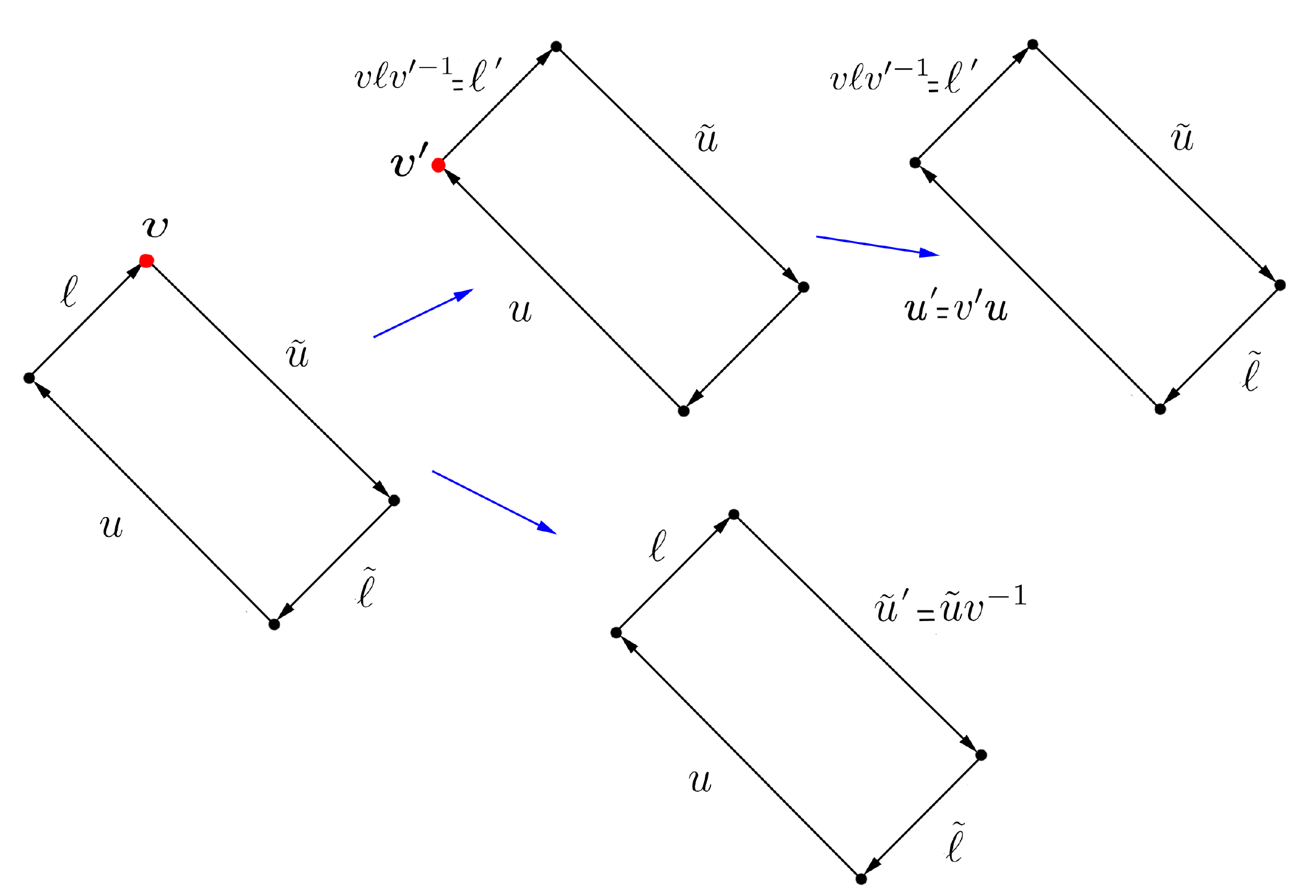}
\caption{The different realizations of the left $\SU(2)$ rotations in terms of ribbons. }\label{leftgauge}
\end{figure}
The $\SU(2)$ holonomies have slightly skewed transformations since they are covariant on one side and braided-covariant on the other:
\be
\left|\begin{array}{l}
u\longrightarrow v'uw^{-1} \\
\tu\longrightarrow w'\tu v^{-1} 
\end{array}\right.
\ee

\medskip

We are now ready to describe how to define the $\SU(2)$ holonomies $u$ and $\tu$ from the spinor variables, using their transformation properties under the $\SU(2)$ action as guide.
First, one must impose a norm-matching condition, $N=\tN$, or equivalently $T_{0}=\tT_{0}$ or in terms of spinors $\la t|t\ra=\la \tt|\tt\ra$.
Starting with $u$, its natural definition is mapping the spinors $|\tt\ra,|\tt]$ to $|\tau],-|\tau\ra$:
\be
u=\f{|\tau]\la \tt|-|\tau\ra[\tt|}{\sqrt{\la \tau|\tau\ra\la \tt|\tt\ra}}\,.
\ee
We first check that this ensures the correct parallel transport:
$$
\f{u\,|\tt\ra\la\tt|\,u^{-1}}{\la \tt|\tt\ra}
\,=\,
\f{|\tau][\tau|}{\la \tau|\tau\ra}
\qquad\Longrightarrow\qquad
u\tT^{-1}u^{-1}=T^{op}\,.
$$
Similarly, we define the $\SU(2)$ holonomy $\tu$ for the tilded sector as the unitary group element mapping the basis $|t\ra,|t]$ to $-|\ttau],|\ttau\ra$:
\be
\tu=\f{|\ttau\ra[t|-|\ttau]\la t|}{\sqrt{\la \ttau|\ttau\ra\la t|t\ra}}\,.
\ee
Once again, this ensures the expected parallel transport:
$$
\f{\tu\,|t\ra\la t|\,\tu^{-1}}{\la t|t\ra}
\,=\,
\f{|\ttau][\ttau|}{\la \ttau|\ttau\ra}
\qquad\Longrightarrow\qquad
\tu T \tu^{-1}=\tT^{op}{}^{-1}\,.
$$
We can further check that $\ell^{-1}\tu^{-1}\tell^{-1}u^{-1}=\id $, by  using the maps between the covariant spinors and dual and their braided-covariant counterparts, and that the straight and tilded sectors indeed define the same $\SL(2,\C)$ group element $D=\ell u = (\tell\tu)^{-1}$
\medskip

Finally, one can compute the Poisson brackets of the components of $u$ and $\tu$, and check that it provides the correct phase space structure.
Let us consider  the $u$ holonomy:
$$
u
\,=\,
\mat{cc}{\alpha & -\bbeta \\ \beta & \balpha}
\,=\,
\f{1}{\sqrt{\la \tau|\tau\ra\la \tt|\tt\ra}}\,
\mat{cc}{-\btau_{1}\btt_{0}+\tau_{0}\tt_{1} & -\btau_{1}\btt_{1}-\tau_{0}\tt_{0}\\ \btau_{0}\btt_{0}+\tau_{1}\tt_{1}&\btau_{0}\btt_{1}-\tau_{1}\tt_{0}}
\,.
$$
In order to compute the Poisson brackets of the components of $u$, we need the Poisson algebra\footnote{
We summarize there all the required Poisson brackets. First, the straight spinors  Poisson-commute with the tilded spinors by definition. 
Then, we recall the brackets for the spinors $t,\tau$ and their complex conjugates:
$$
\{t_{0},t_{1}\}=\f{i\ka}{2}t_{0}t_{1},\qquad
\{\bt_{0},\bt_{1}\}=-\f{i\ka}{2}\bt_{0}\bt_{1},\qquad
\{t_{0},\bt_{1}\}=\{\bt_{0},t_{1}\}=0,
$$
$$
\{\bt_{0},t_{0}\}=-\f{i\ka}{2}|t_{0}|^{2} +i\ka e^{\f{\ka}2 N},
\qquad
\{\bt_{1},t_{1}\}=\f{i\ka}{2}|t_{1}|^{2} +i\ka e^{-\f{\ka}2 N}\,,
$$
$$
\{\tau_{0},\tau_{1}\}=-\f{i\ka}{2}\tau_{0}\tau_{1},\qquad
\{\btau_{0},\btau_{1}\}=\f{i\ka}{2}\btau_{0}\btau_{1},\qquad
\{\tau_{0},\btau_{1}\}=-\f{i\ka}{2}\tau_{0}\btau_{1},\qquad
\{\btau_{0},\tau_{1}\}=\f{i\ka}{2}\btau_{0}\tau_{1},
$$
$$
\{\btau_{0},\tau_{0}\}={+i\ka}\,e^{{\ka} N_{0}},
\qquad
\{\btau_{1},\tau_{1}\}={+i\ka}\,e^{{\ka} N}\,.
$$
Finally, the contribution of the norm-factors is computed using the explicit expression of the norms $\la t|t\ra=(e^{\f{\ka N}{2}}-e^{\f{-\ka N}{2}})$ and $\la \tau|\tau\ra=(e^{{\ka N}}-1)$:
\bes 
&&
\{\la t|t\ra^{-\f12} , t_{A}\}
\,=\,
-\f{i\ka}{4}\la t|t\ra^{-\f32}\,(e^{\f\ka2 N}+e^{-\f\ka2 N})\,t_{A},
\qquad
\{\la t|t\ra^{-\f12} , \bt_{A}\}
\,=\,
\f{i\ka}{4}\la t|t\ra^{-\f32}\,(e^{\f\ka2 N}+e^{-\f\ka2 N})\,\bt_{A},
\nn\\ &&
\{\la \tau|\tau\ra^{-\f12} , \tau_{A}\}
\,=\,
-\f{i\ka}{2}\la \tau|\tau\ra^{-\f32}\,e^{{\ka N}}\,\tau_{A},
\qquad
\{\la \tau|\tau\ra^{-\f12} , \btau_{A}\}
\,=\,
\f{i\ka}{2}\la \tau|\tau\ra^{-\f32}\,e^{{\ka N}}\,\btau_{A}\,.\nn
\ees
}
 of the spinor components given earlier in \eqref{t com},   \eqref{tau com} and \eqref{cross}.
Being careful to the norm factors, which are actually rather important and not to be neglected, we compute the Poisson brackets of $u$ with itself and find:
\be
\{\alpha,\beta\}\simeq\,-\f{i\ka}{2}\alpha\beta\,,\quad
\{\alpha,\bbeta\}\simeq\,-\f{i\ka}{2}\alpha\bbeta\,,\quad
\{\alpha,\balpha\}\simeq\,i\ka\,|\beta|^2,\quad
\{\beta,\bbeta\}\simeq\,0\,,
\ee
which are weak equalities up to assuming the norm-matching constraint $N-\tN=0$. This is  valid since all these variables $\ell, \tell, u, \tu$ have a vanishing Poisson bracket with that constraint.
These are the expected Poisson brackets for the $\SU(2)$ holonomy $u$ from the $\SL(2,\C)$ phase space formulas. Proceeding similarly, it is straightforward to check that we recover the whole phase space with the correct brackets between the left decomposition variables, between the right decomposition variables (i.e. the tilded variables) as well   between these two sets of variables \eqref{cross}. 

We have therefore the following theorem.

\begin{theorem}
Consider $\cS_\ka\sim \C^2$  the phase space given in terms of the following (non-canonical) Poisson brackets ($t_A\in \C^2$)
\bes
\{\bt_{0},t_{0}\}=\f{i\ka}{2}(-|t_{0}|^{2} +2e^{\f{\ka}2 N}),
\quad
\{\bt_{1},t_{1}\}=\f{i\ka}{2}(|t_{1}|^{2} +2 e^{-\f{\ka}2 N})\,,\nn
\\ 
\{t_{0},t_{1}\}=\f{i\ka}{2}t_{0}t_{1},\quad
\{\bt_{0},\bt_{1}\}=-\f{i\ka}{2}\bt_{0}\bt_{1},\quad
\{t_{0},\bt_{1}\}=\{\bt_{0},t_{1}\}=0\,, \nn
\ees
and its subspace $\cS_{\ka*}\equiv \C^2/ \{t_a\in\C^2, \, \la t|t\ra=0\}$. 

The symplectic reduction $\cP_\ka= (\cS_{\ka*}\times\cS_{\ka*})//\cM$  of the phase space $\cS_{\ka*}\times\cS_{\ka*}$ (parametrized by $(t_A,\tilde t_A)$)
by the matching norm condition $\cM=\la t|t\ra-\la \tilde t|\tilde t\ra=0$ is isomorphic to the phase space $\SL(2,\C)$ which Poisson structure is  
$$
\{D_{1},D_{2}\}=-rD_{1}D_{2}-D_{1}D_{2}r^\dagger\,, \quad D\in\SL(2,\C), \qquad
r=\f\ka4\sum_i \tau_i\ot \sigma_i,
\quad \tau_{i}=i(\sigma_{i}-\f12[\sigma_{3},\sigma_{i}]).
$$
\end{theorem}
This theorem is the classical analogue of the Schwinger-Jordan representation of  $\UQ$ in terms of harmonic oscillators \cite{mac, bied}. Note furthermore that we can relate $\cS_\ka$ to the undeformed phase space $\C^2$, as we have recalled in \eqref{ho def}. Hence there is a symplectomorphism
between the undeformed phase space $\C^2$ and $\cS_\ka$. This symplectomorphism can then be naturally extended between the spaces $\cP= (\C^2_{*}\times\C^2_{*})//\cM$ and $\cP_\ka= (\cS_{\ka*}\times\cS_{\ka*})//\cM$. We recall that $\cP$ is the phase space behind the (flat) twisted geometries construction and is such that $\cP\sim T^*\SU(2)$ \cite{twisted1, twisted-rev}. Hence we have  recovered in a new manner the well-known fact that $T^*\SU(2)$ and $\SL(2,\C)$ are symplectomorphic as phase spaces (while still obviously different as groups).

\section{Towards hyperbolic twisted geometries}

We would like to use the spinor phase space developed above to parametrize the hyperbolic deformation of twisted geometries for loop gravity. We consider a graph, with oriented edges and chosen order around each vertex (i.e all the attached edges are ordered). We attach spinor variables to each half-edge: around each vertex ${\vt}$, for each attached edge $e\ni {\vt}$, we define a spinor $\zeta_{e,{\vt}}\in\C^{2}$ with canonical Poisson brackets and then as developed previously spinors $\zeta_{e,\vt}^{\ka}$, $t_{e,{\vt}}$ and $\tau_{e,{\vt}}$ satisfying deformed Poisson brackets. Then for each edge we have a pair of spinors, one at each end. We naturally associate the source vertex to the straight sector and the target vertex to the tilded sector: we write for short $\zeta_{e}=\zeta_{e,s(e)}$ at the target vertex ${\vt}=s(e)$ and $\tzeta_{e}=\zeta_{e,t(e)}$ at the source vertex ${\vt}=t(e)$.
Assuming norm-matching conditions along the edges,
$$
M_{e}=\la \zeta_{e,s}|\zeta_{e,s}\ra-\la \zeta_{e,t}|\zeta_{e,t}\ra
=\la \zeta_{e}|\zeta_{e}\ra-\la \tzeta_{e}| \tzeta_{e}\ra=0\,,
$$
we can define all the group elements $\ell_{e},\tell_{e}\in\SB(2,\C)$ and $u_{e},\tu_{e}\in\SU(2)$ along the edges. All of these variables have vanishing Poisson brackets with the norm-matching constraint. The norm-matching constraint generates a $\U(1)$ action on the spinors, multiplying $\zeta_{e}$ and $\tzeta_{e}$ by opposite phases. The group elements $\ell$'s and $u$'s are all invariant under such phase shifts. And this leads to one copy of the $\SL(2,\C)$ phase space on each edge with the group element $D_{e}=\ell_{e}u_{e}=(\tell_{e}\tu_{e})^{-1}$ and Poisson brackets given in terms of the classical r-matrix.

Up to now, we have simply introduce a re-parametrization of the standard phase space $(\C^{2})^{2E}//\U(1)^{E}$, where we are taking the symplectic quotient by the matching conditions along every edge, in terms of the non-trivial and non-canonical variables $\ell_{e},\tell_{e},u_{e},\tu_{e}$. The big difference with the standard spinor networks and twisted geometries is the deformed action of the $\SU(2)$ gauge group at the vertices. Indeed the initial canonical spinors $\zeta_{e},\tzeta_{e}$ do not transform simply as spin-1/2 vectors under $\SU(2)$ transformations. It is instead the non-canonical spinors $t_{e},\tt_{e}$ that transform linearly under   the action of $\SU(2)$.
This $\SU(2)$ action can be in principle transposed on the canonical spinors $\zeta_{e},\tzeta_{e}$, it is highly non-linear and the explicit expression is rather cumbersome.
Taking this into account, it is not obvious that the matching conditions $M_{e}=\la \zeta_{e}|\zeta_{e}\ra-\la \tzeta_{e}| \tzeta_{e}\ra=0$ are invariant under the $\SU(2)$ action. There are nevertheless exactly equivalent to the obviously $\SU(2)$-invariant constraint written in terms of the non-canonical spinors $t$'s,
$$
\cM_{e}=
\la t_{e}|t_{e}\ra-\la \tt_{e}| \tt_{e}\ra=0
\,.
$$

\begin{figure}
\includegraphics[height=45mm]{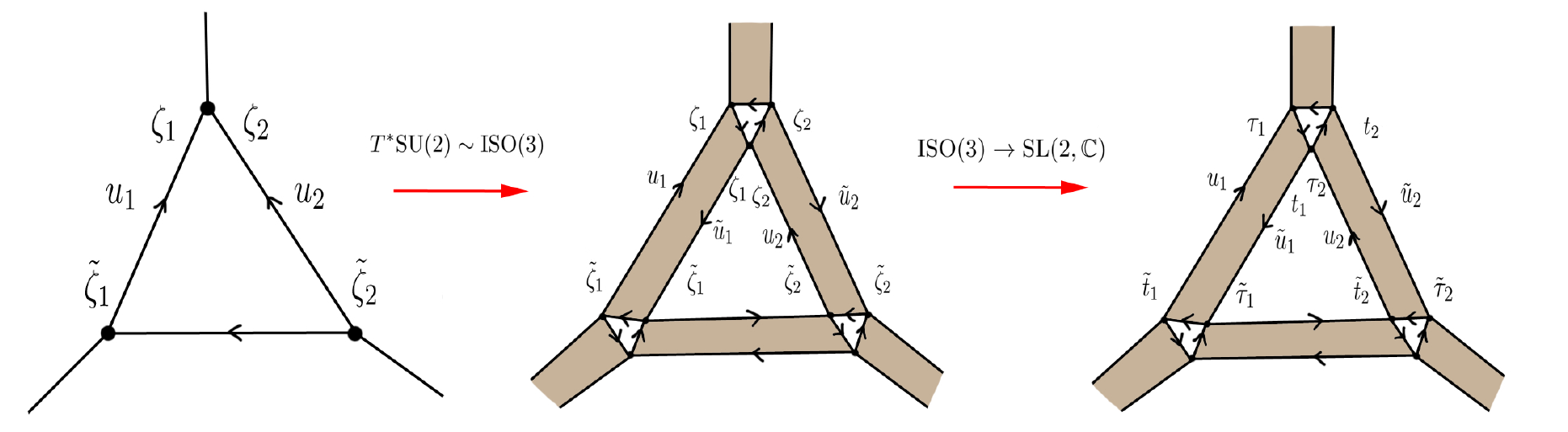}
\caption{On the left side, we have a standard spinor network. We flatten the graph to encode $T^*\SU(2)$ as $\ISO(3)$. In this case, $\tu_i=u\mone_i$, and we have the same spinors standing at the extremities of each ribbon. On the right side, we deform $\ISO(3)$ into $\SL(2,\C)$. Hence the braided covariant spinors appear: the extremities of the ribbon have both a covariant and a braided covariant spinor. }
\label{spinornetwork}
\end{figure}

Actually the new setting is even more subtle. In order to define the $\SU(2)$ action at a vertex ${\vt}$, we need to consider the chosen ordering of the edges around it. Let's number these edges $e_{i}$ with $i=1..n$. As shown and explained in \cite{HyperbolicPhaseSpace}, assuming that all the edges are oriented outward, the $\SU(2)$ transformations are generated by the deformed closure  constraints (or Gauss law in the context of loop gravity) $\cG_{\vt}=\ell_{1}..\ell_{n}=\id$. The product is obviously non-abelian and the ordering of the triangular matrices matters. More precisely, the $\SU(2)$ transformation for one (half-)edge $e_{i}$ will get braided by all the previous $\ell_{k}$ for $k<i$. This reads for a transformation $v\in\SU(2)$;
\be
\left|\begin{array}{rcr}
\ell_{1}
& \quad \longrightarrow \quad &
v\ell_{1}v_{1}^{-1} \\
\ell_{2}
& \quad \longrightarrow \quad &
v_{1}\ell_{2}v_{2}^{-1} \\
\ell_{i}
& \quad \longrightarrow \quad &
v_{i-1}\ell_{i}v_{i}^{-1} \\
\ell_{n}
& \quad \longrightarrow \quad &
v_{n-1}\ell_{n}v^{-1} \\
\end{array} \right.\,\,,
\ee
where the last $v'$ factor is equal to the initial $\SU(2)$ transformation $v_{n}=v$ due to the condition $\ell_{1}..\ell_{n}=\id$ itself.
As shown in \cite{HyperbolicPhaseSpace}, these $\SU(2)$ transformations are Poisson-generated by the Hermitian square of the closure constraints $\cG_{\vt}^\dagger\cG_{\vt}$ as:
\be
\delta_{v}f = -\ka^{-1\,}\prod_{k=1}^n \lambda_{k}^{-2}\,\{ \tr V \cG_{\vt}^\dagger\cG_{\vt}\,,\,f\}
\qquad\textrm{with}\quad
v\sim\id + i \eps\cdot\vsigma
=
\id + i \mat{cc}{\eps_{z} & \eps_{-} \\ \eps_{+} & \eps_{z}} 
,\quad
V=\mat{cc}{2\eps_{z} & \eps_{-} \\ \eps_{+} & 0} 
\,,
\ee
for infinitesimal transformation $v\sim\id$ with $\eps\sim 0$. For $n=1$, we recover the infinitesimal transformations already given in \eqref{su2transfneq1}.
If an edge $e_{k}$ is incoming instead of outgoing, one simply replaces $\ell_{k}$ by $\tell_{k}$ in the closure constraint and nothing else changes.
The interested reader will find detailed explanations and proofs for this in \cite{HyperbolicPhaseSpace}.
We can summarize all these structures by a graphical representation of our deformed spin network as ribbon graphs, as illustrated on Fig. \ref{spinornetwork}.

Moreover, the previous work \cite{HyperbolicPhaseSpace} further provided a geometrical interpretation for the closure constraints at least in a (2+1)-dimensional setting. Considering trivalent graphs, one can map uniquely a triplet of triangular matrices satisfying the closure $\ell_{1}\ell_{2}\ell_{3}=\id$ onto a hyperbolic triangle (up to translations) such that the $\ell$'s are interpreted as the ``hyperbolic normals'' to the triangle's edges. Then, assuming a ($\SU(2)$-invariant) flatness condition around every loop (in terms of the $\SU(2)$ holonomies $u_{e}$ and $\tu_{e}$), one could glue these hyperbolic triangles consistently together within the 3-hyperboloid (of unit time-like vectors in the 4d Minkowsi space-time $\R^{4}$). However, a (3+1)-d geometric interpretation of the closure constraints, and thus of these deformed spinor networks, in terms of hyperbolic tetrahedra and polyhedra has not yet been worked out. This is a highly non-trivial mathematical issue and we leave for future investigation. This would provide us with hyperbolic twisted geometries as a generalization to non-vanishing cosmological constant  of the standard twisted geometries describing the kinematical configurations of loop gravity.

\section*{Discussion}

We have shortly reviewed the deformed phase space for loop gravity and the associated deformed action of $\SU(2)$ introduced in \cite{HyperbolicPhaseSpace} in order to represent discrete hyperbolic surfaces and to yield directly spin networks for $\cU_{q}(\su(2))$ upon canonical quantization \cite{ham}. Here, we have introduced spinor variables with canonical Poisson brackets as Darboux coordinates for this phase space and we have defined non-canonical spinor variables that allow to represent the deformed phase space on a given graph as a spinor network. The resulting structures on the graph are a pair of spinors on each edge, one attached to each of its ends, provided with norm-matching constraints for every edge and deformed closure constraints for every vertex. This ensures that one can reconstruct the full $\SL(2,\C)$ phase space on each edge, given in terms of a group element $D_{e}=\ell_{e}u_{e}\in\SL(2,\C)\sim\SB(2,\C)\bowtie \SU(2)$ together with a Poisson bracket defined in terms of the classical $r$-matrix of $\sl_{2}$, and that we have a local $\SU(2)$ invariance at each vertex.

Just like the use of spinor variables for twisted geometries in standard loop quantum gravity (e.g. \cite{un-spinor,twisted-rev}), we hope that this spinorial formalism for the hyperbolically-deformed loop gravity phase space will similarly shed light on the construction of observables and the interpretation of the deformed phase space as 2d and 3d hyperbolic geometries, at the classical level. We expect that it should provide some insights on the definition of  coherent states at the quantum level. This work should be considered as setting up the framework for a larger program whose next steps would be:
\begin{itemize}

\item Identify the $\SU(2)$-invariant observables at the vertices in terms of suitable scalar products between spinors and investigate the  deformed $\u(N)$ structure of their Poisson algebra. This should be the classical analogue of the $\cU_q(\u(N))$ structure found in the quantum case \cite{ours2}.

\item Build on the work \cite{ours1,ours2}  and 
investigate the relation between the classical flatness constraints and the  spinfoam dynamics for 3d quantum gravity with negative cosmological constant through the analysis of recursion relations for the $q$-deformed $6j$-symbols using spinor operators, as done in \cite{Bonzom:2011nv}.

\item Study the $\SU(2)$ flatness holonomy constraint defined as $\overrightarrow{\prod} u_{e}=\id$ in the previous work \cite{HyperbolicPhaseSpace}, analyze its expansion in terms of the deformation parameter $\ka$ and understand the deviations from the standard flat spinor network case  and the origin of  the (hyperbolic) curvature in the continuum limit.

\item Extend the general construction to the spherical case, where quasi-Poisson Lie group structures will likely  be the right tool \cite{spherical}.

\item Understand the geometrical meaning of the deformed spinor network in terms of 3d geometry embedded in a 3+1d space-time and extend the construction to a deformed twistor phase space and appropriate simplicity constraints as done in the standard flat case \cite{Livine:2011vk,Speziale:2012nu}.

\end{itemize}

At the end of the day, we believe  spinorial variables to be a very promising and relevant tool to investigate the quantization of loop gravity with a non-vanishing cosmological constant and its relation to the quantum deformation of the gauge group.

\section*{Acknowledgement}
M. Dupuis and F. Girelli acknowledge financial support from the Government of Canada through
respectively a Banting fellowship and a NSERC Discovery grant.


\bibliographystyle{bib-style}
\bibliography{qspace}

\providecommand{\href}[2]{#2}\begingroup\raggedright\begin{thebibliography}{10}

\bibitem{Schroers:2011wn}
B.~J. Schroers, ``{Quantum gravity and non-commutative spacetimes in three
  dimensions: a unified approach},'' Acta Phys.Polon.Supp. {\bf 4} (2011)
  379--402,
\href{http://arXiv.org/abs/1105.3945}{{\texttt{arXiv:1105.3945}}}.

\bibitem{Witten1}
E.~Witten, ``{(2+1)-Dimensional Gravity as an Exactly Soluble System},''
  Nucl.Phys. {\bf B311} (1988)
46.

\bibitem{Witten2}
E.~Witten, ``{Quantum Field Theory and the Jones Polynomial},''
  Commun.Math.Phys. {\bf 121} (1989)
351.

\bibitem{Alekseev:1994pa}
A.~Y. Alekseev, H.~Grosse, and V.~Schomerus, ``{Combinatorial quantization of
  the Hamiltonian Chern-Simons theory},'' Commun.Math.Phys. {\bf 172} (1995)
  317--358,
\href{http://arXiv.org/abs/hep-th/9403066}{{\texttt{arXiv:hep-th/9403066}}}.

\bibitem{Buffenoir:2002tx}
E.~Buffenoir, K.~Noui, and P.~Roche, ``{Hamiltonian quantization of
  Chern-Simons theory with SL(2,C) group},'' Class.Quant.Grav. {\bf 19} (2002)
  4953,
\href{http://arXiv.org/abs/hep-th/0202121}{{\texttt{arXiv:hep-th/0202121}}}.

\bibitem{Meusburger:2008bs}
C.~Meusburger and K.~Noui, ``{The Hilbert space of 3d gravity: quantum group
  symmetries and observables},'' Adv.Theor.Math.Phys. {\bf 14} (2010)
  1651--1716,
\href{http://arXiv.org/abs/0809.2875}{{\texttt{arXiv:0809.2875}}}.

\bibitem{cat}
C.~Meusburger and B.~Schroers, ``{Quaternionic and Poisson-Lie structures in 3d
  gravity: The Cosmological constant as deformation parameter},'' J.Math.Phys.
  {\bf 49} (2008) 083510,
\href{http://arXiv.org/abs/0708.1507}{{\texttt{arXiv:0708.1507}}}.

\bibitem{Turaev:1992hq}
V.~Turaev and O.~Viro, ``{State sum invariants of 3 manifolds and quantum 6j
  symbols},'' Topology {\bf 31} (1992)
865--902.

\bibitem{Mizoguchi:1991hk}
S.~Mizoguchi and T.~Tada, ``{Three-dimensional gravity from the Turaev-Viro
  invariant},'' Phys.Rev.Lett. {\bf 68} (1992) 1795--1798,
\href{http://arXiv.org/abs/hep-th/9110057}{{\texttt{arXiv:hep-th/9110057}}}.

\bibitem{woodward}
Y.~U. Taylor and C.~T. Woodward, ``{6j symbols for $U_q
  (\mathfrak{s}\mathfrak{l}_2 )$ and non-Euclidean tetrahedra},'' Selecta
  Mathematica {\bf 11} (2005) 539--571,
  \href{http://arXiv.org/abs/math/0305113}{{\texttt{arXiv:math/0305113}}}.

\bibitem{Freidel:2004nb}
L.~Freidel and D.~Louapre, ``{Ponzano-Regge model revisited II: Equivalence
  with Chern-Simons},''
\href{http://arXiv.org/abs/gr-qc/0410141}{{\texttt{arXiv:gr-qc/0410141}}}.

\bibitem{ours1}
M.~Dupuis and F.~Girelli, ``{Quantum hyperbolic geometry in loop quantum
  gravity with cosmological constant},'' Phys.Rev. {\bf D87} (2013), no.~12,
  121502, \href{http://arXiv.org/abs/1307.5461}{{\texttt{arXiv:1307.5461}}}.

\bibitem{ours2}
M.~Dupuis and F.~Girelli, ``{Observables in Loop Quantum Gravity with a
  cosmological constant},''
  \href{http://arXiv.org/abs/1311.6841}{{\texttt{arXiv:1311.6841}}}.

\bibitem{KauffmanBracketLQG}
K.~Noui, A.~Perez, and D.~Pranzetti, ``{Non-commutative holonomies in 2+1 LQG
  and Kauffman's brackets},'' J.Phys.Conf.Ser. {\bf 360} (2012) 012040,
  \href{http://arXiv.org/abs/1112.1825}{{\texttt{arXiv:1112.1825}}}.

\bibitem{pranzetti}
D.~Pranzetti, ``{Turaev-Viro amplitudes from 2+1 Loop Quantum Gravity},''
  \href{http://arXiv.org/abs/1402.2384}{{\texttt{arXiv:1402.2384}}}.

\bibitem{HyperbolicPhaseSpace}
V.~Bonzom, M.~Dupuis, F.~Girelli, and E.~R. Livine, ``{Deformed phase space for
  3d loop gravity and hyperbolic discrete geometries},''
  \href{http://arXiv.org/abs/1402.2323}{{\texttt{arXiv:1402.2323}}}.

\bibitem{ham}
V.~Bonzom, M.~Dupuis, and F.~Girelli, ``{Towards the Turaev-Viro amplitudes
  from a Hamiltonian constraint},''
  \href{http://arXiv.org/abs/1403.7121}{{\texttt{arXiv:1403.7121}}}.

\bibitem{twisted1}
L.~Freidel and S.~Speziale, ``{Twisted geometries: A geometric parametrisation
  of SU(2) phase space},'' Phys.Rev. {\bf D82} (2010) 084040,
  \href{http://arXiv.org/abs/1001.2748}{{\texttt{arXiv:1001.2748}}}.

\bibitem{twisted2}
L.~Freidel and S.~Speziale, ``{From twistors to twisted geometries},''
  Phys.Rev. {\bf D82} (2010) 084041,
\href{http://arXiv.org/abs/1006.0199}{{\texttt{arXiv:1006.0199}}}.

\bibitem{twisted-rev}
M.~Dupuis, S.~Speziale, and J.~Tambornino, ``{Spinors and Twistors in Loop
  Gravity and Spin Foams},'' PoS {\bf QGQGS2011} (2011) 021,
  \href{http://arXiv.org/abs/1201.2120}{{\texttt{arXiv:1201.2120}}}.

\bibitem{un-spinor}
E.~F. Borja, L.~Freidel, I.~Garay, and E.~R. Livine, ``{U(N) tools for Loop
  Quantum Gravity: The Return of the Spinor},'' Class.Quant.Grav. {\bf 28}
  (2011) 055005,
\href{http://arXiv.org/abs/1010.5451}{{\texttt{arXiv:1010.5451}}}.

\bibitem{spinor-lqg}
E.~R. Livine and J.~Tambornino, ``{Spinor Representation for Loop Quantum
  Gravity},'' J.Math.Phys. {\bf 53} (2012) 012503,
\href{http://arXiv.org/abs/1105.3385}{{\texttt{arXiv:1105.3385}}}.

\bibitem{spinor-holonomy}
E.~R. Livine and J.~Tambornino, ``{Holonomy Operator and Quantization
  Ambiguities on Spinor Space},'' Phys.Rev. {\bf D87} (2013), no.~10, 104014,
\href{http://arXiv.org/abs/1302.7142}{{\texttt{arXiv:1302.7142}}}.

\bibitem{spinor-coh1}
M.~Dupuis and E.~R. Livine, ``{Revisiting the Simplicity Constraints and
  Coherent Intertwiners},'' Class.Quant.Grav. {\bf 28} (2011) 085001,
\href{http://arXiv.org/abs/1006.5666}{{\texttt{arXiv:1006.5666}}}.

\bibitem{spinor-coh2}
M.~Dupuis and E.~R. Livine, ``{Holomorphic Simplicity Constraints for 4d
  Spinfoam Models},'' Class.Quant.Grav. {\bf 28} (2011) 215022,
\href{http://arXiv.org/abs/1104.3683}{{\texttt{arXiv:1104.3683}}}.

\bibitem{Freidel:2012ji}
L.~Freidel and J.~Hnybida, ``{On the exact evaluation of spin networks},''
\href{http://arXiv.org/abs/1201.3613}{{\texttt{arXiv:1201.3613}}}.

\bibitem{Bonzom:2011nv}
V.~Bonzom and E.~R. Livine, ``{A New Hamiltonian for the Topological BF phase
  with spinor networks},'' J.Math.Phys. {\bf 53} (2012) 072201,
\href{http://arXiv.org/abs/1110.3272}{{\texttt{arXiv:1110.3272}}}.

\bibitem{Livine:2011up}
E.~R. Livine and M.~Martin-Benito, ``{Classical Setting and Effective Dynamics
  for Spinfoam Cosmology},'' Class.Quant.Grav. {\bf 30} (2013) 035006,
\href{http://arXiv.org/abs/1111.2867}{{\texttt{arXiv:1111.2867}}}.

\bibitem{Bonzom:2012bn}
V.~Bonzom and E.~R. Livine, ``{Generating Functions for Coherent
  Intertwiners},'' Class.Quant.Grav. {\bf 30} (2013) 055018,
\href{http://arXiv.org/abs/1205.5677}{{\texttt{arXiv:1205.5677}}}.

\bibitem{francesco}
F.~Costantino and J.~Marche, ``Generating series and asymptotics of classical
  spin networks,''
  \href{http://arXiv.org/abs/1103.5644}{{\texttt{arXiv:1103.5644}}}.

\bibitem{mac}
A.~Macfarlane, ``{On q Analogs of the Quantum Harmonic Oscillator and the
  Quantum Group SU(2)-q},'' J.Phys. {\bf A22} (1989)
4581.

\bibitem{bied}
L.~Biedenharn, ``{The Quantum Group SU(2)-q and a q Analog of the Boson
  Operators},'' J.Phys. {\bf A22} (1989)
L873.

\bibitem{spherical}
T.~Treloar, ``{The symplectic geometry of polygons in the 3-sphere},'' arXiv
  preprint math/0009193 (2000).

\bibitem{Livine:2011vk}
E.~R. Livine, S.~Speziale, and J.~Tambornino, ``{Twistor Networks and Covariant
  Twisted Geometries},'' Phys.Rev. {\bf D85} (2012) 064002,
\href{http://arXiv.org/abs/1108.0369}{{\texttt{arXiv:1108.0369}}}.

\bibitem{Speziale:2012nu}
S.~Speziale and W.~M. Wieland, ``{The twistorial structure of loop-gravity
  transition amplitudes},'' Phys.Rev. {\bf D86} (2012) 124023,
\href{http://arXiv.org/abs/1207.6348}{{\texttt{arXiv:1207.6348}}}.

\end{thebibliography}\endgroup

\end{document}